# Spin structure and magnetic phase transitions in TbBaCo$_2$O$_{5.5}$

V. P. Plakhty,[1] Yu. P. Chernenkov,[1*] S. N. Barilo,[2] A. Podlesnyak,[3] E. Pomjakushina,[3]
D. D. Khalyavin,[2] E. V. Moskvin,[1] S.V. Gavrilov[1]

[1]*Petersburg Nuclear Physics Institute, Gatchina, 188300 St. Petersburg, Russia*
[2]*Institute of Solid State & Semiconductor Physics, National Acad. Sci., Minsk 220072, Belarus*
[3]*Laboratory for Neutron Scattering, ETH Zürich & Paul Scherrer Institut CH-5232 Villigen PSI, Switzerland*

Spin ordering in TbBaCo$_2$O$_{5.5}$ and its temperature transformation reproducible for differently synthesized samples are studied. First of all, the polymorphism due to the oxygen ordering with the average content close to 5.5 is investigated. One of ceramic samples (I), in addition to the main phase $a_p \times 2a_p \times 2a_p$, $Pmmm$ ($Z = 2$), contained about 25% of the phase $a_p \times a_p \times 2a_p$, $Pmmm$, ($Z = 1$) with statistical distribution of oxygen over the apical sites, where $a_p$ is parameter of perovskite cell. The other sample (II) contained a single phase $a_p \times 2a_p \times 2a_p$, $Pmmm$ ($Z = 2$) with well defined octahedral and pyramidal sublattices. Treatment of neutron diffraction patterns of the sample I itself gives a sophisticated spin structure. Knowing the structure of sample II, one can chose only proper magnetic lines, which give exactly the same results as for sample II. Above the Néel temperature $T_N \approx 290$ K, there is a structural transition to the phase $2a_p \times 2a_p \times 2a_p$, $Pmma$. At $T_N$, the spins order with the wave vector $\mathbf{k}_{19} = 0$ (phase 1). At $T_1 \approx 255$ K, a magnetic transition takes place to the phase 2 with $\mathbf{k}_{22} = \mathbf{b}_3/2$. At $T_2 \approx 170$ K, the crystal structure changes to $2a_p \times 2a_p \times 4a_p$, $Pcca$ ($Z = 4$). The wave vector of the spin structure becomes again $\mathbf{k}_{19} = 0$ (phase 3). The basis functions of irreducible representations of the group $G_\mathbf{k}$ have been found. Using results of this analysis, the magnetic structure in all phases is determined. The spins are always parallel to the **x** axis, and the difference is in the values and mutual orientation of the moments in the ordered non-equivalent pyramidal or octahedral positions. Spontaneous moment $M_0 = 0.30(3)$ $\mu_B$/Co at $T = 260$ K is due to ferrimagnetic ordering of the moments $M_{Py1} = 0.46(9)$ $\mu_B$ and $M_{Py2} = -1.65(9)$ $\mu_B$ in pyramidal sites (Dzyaloshinskii-Moriya canting is forbidden by symmetry). The moments in the non-equivalent octahedral sites are: $M_{Oc1} = -0.36(9)$ $\mu_B$, $M_{Oc2} = 0.39(9)$ $\mu_B$. At $T = 230$ K, $M_{Py1} = 0.28(8)$ $\mu_B$, $M_{Py2} = 1.22(8)$ $\mu_B$, $M_{Oc1} = 1.39(8)$ $\mu_B$, $M_{Oc2} = -1.52(8)$ $\mu_B$. At $T = 100$ K, $M_{Py1} = 1.76(6)$ $\mu_B$, $M_{Py2} = -1.76$ $\mu_B$, $M_{Oc1} = 3.41(8)$ $\mu_B$, $M_{Oc2} = -1.47(8)$ $\mu_B$. The moment values together with the ligand displacements are used to analyze the picture of spin-state/orbital ordering in each phase.

PACS numbers: 61.12.Ld, 61.66.Fn, 71.30.+h, 75.25.+z

## I. INTRODUCTION

The giant magnetoresistive cobaltites $R$BaCo$_2$O$_{5.5}$ ($R$ = Y, or rare earth) demonstrate spin ordering, apparently, coupled with the orbital ordering on the background of Co$^{3+}$ spin-state transitions. As a result of competition between the intra-atomic exchange and the crystal field, Co$^{3+}$ ions can exist in the low-spin state (LS, $t_{2g}^6 e_g^0$, $S = 0$), intermediate-spin state (IS, $t_{2g}^5 e_g^1$, $S = 1$) and the high-spin state (HS, $t_{2g}^4 e_g^2$, $S = 2$).[1] Due to rather small energy gaps of about 30 ÷ 100 meV between these states,[2,3] the LS evolves to the IS and then to the HS as the temperature increases. These temperature-dependent spin-state transformations may be accompanied by both the spin structure and the spin-state ordering transitions, which were investigated using neutron diffraction[4-6] and magnetization measurements.[7]

The crystal structure shown in Fig. 1(a) is found now for all of the series members.[8-11] The space group $D_{2h}^1$–$Pmmm$ is usually considered, and the unit cell parameters are expressed through parameter $a_p$ of the pseudocubic perovskite cell as $a_1 \approx a_p$, $a_2 \approx 2a_p$, $a_3 \approx 2a_p$. The structure motif is given by the oxygen octahedra and square pyramids, which coordinate the Co$^{3+}$ ions and alternate along **y**–axis. Apical oxygen ions shown by the open circles can be partially occupied. This means that even at the average oxygen content of 5.5 some disorder can appear due to the oxygen redistribution among these sites. The regular study[12] of the crystal lattice dependence on the oxygen content from 5.00 to 5.52 in YBaCo$_2$O$_{5+x}$ had discovered, in addition to $a_p \times 2a_p \times 2a_p$, a superstructure with tetragonal unit cell $3a_p \times 3a_p \times 2a_p$ at $0.25 \le x \le 0.44$. At low oxygen content, $x \le 0.19$, nearly tetragonal unit cell $a_p \times a_p \times 2a_p$ shown in Fig. 1(b) was discovered. Similar structure was observed at $x = 0.4$ for a fast-cooled sample.[4] A variety of different crystal phases[12] inevitably leads to controversies of the magnetic structures determined up to now.

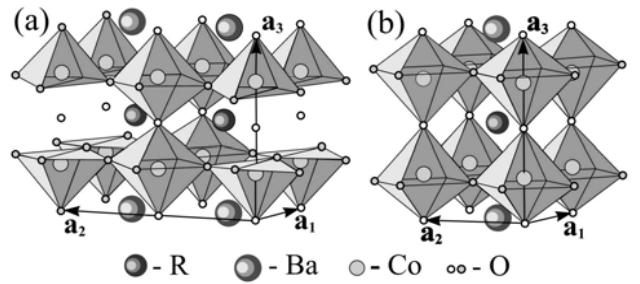

FIG. 1. (a) Unit cell of the layered perovskite $R$BaCo$_2$O$_{5.5}$. Open small circles show the crystallographic positions, which can be partially occupied by the O$^{2-}$ ions. (b) Unit cell $a_p \times a_p \times 2a_p$ that has been found at statistical occupancies of the apical oxygen sites.

The first neutron diffraction study of magnetic ordering[4] was carried out on the fast cooled ceramic of TbBaCo$_2$O$_{5.4}$ with the crystal structure described in Fig. 1(b). The magnetic unit cell at $T$ = 150 K was determined as twice enlarged chemical one in all three directions. All six nearest neighbors of any spin had an opposite direction. Magnetic moments had two components, $M_y$ = 1.33(3) $\mu_B$ and $M_z$ = 0.70(5) $\mu_B$. This sample had no phase with spontaneous moment at about 260 ÷ 270 K. Unfortunately, no neutron diffraction data were presented for a slowly cooled sample, which would have the crystal structure of interest, described in Fig. 1(a).

Neutron diffraction studies[5] that have been performed on a powder sample of NdBaCo$_2$O$_{5.47}$ show two phase transitions, to the antiferromagnetic phase at $T_N$ ~ 275 K and to the spin-state ordered (SSO) phase at $T_{SSO}$ ~ 230 K. The spin-state ordering is discovered in octahedra with the moments at 85 K of 2.6(2) $\mu_B$ and 0.1(1) $\mu_B$, which are interpreted as IS and LS, respectively. The cobalt ions of the same spin state are arranged in the rows along [0,0,1], which alternate along [1,0,0]. Some widening of magnetic reflections in comparison with the nuclear ones is explained by a finite correlation length $\xi$ ~ 350 Å of the SSO. Nothing special has been detected in the temperature range of the spontaneous moment, and there is no explanation of its nature. It is mentioned that the same behavior has been observed in TbBaCo$_2$O$_{5.46}$ ($T_N$ ~ 295 K, $T_{SSO}$ ~ 250 K). On the other hand no SSO has been found in NdBaCo$_2$O$_{5.38}$, which indicates that this state exists in a narrow range of the oxygen content near $x$ = 0.5.

The third and, to our knowledge, the last neutron-diffraction work[6] has been made on a single crystal of TbBaCo$_2$O$_{5.52(2)}$ grown by a floating zone method. As-grown crystal was annealed in flowing oxygen. A number of superlattice reflections have been detected with the wave vectors $\mathbf{k}_7 = \mathbf{b}_1/3$ (or $\mathbf{k}_8 = 2\mathbf{b}_2/3$), $\mathbf{k}_7 = \mathbf{b}_1/6$ (or $\mathbf{k}_8 = \mathbf{b}_2/3$) $\mathbf{k}_9 = \mathbf{b}_3/4$ in addition to $\mathbf{k}_{22} = \mathbf{b}_3/2$. Here the wave vectors $\mathbf{k}_i$ are given in Kovalev's notation,[13] and $\mathbf{b}_1$, $\mathbf{b}_2$, $\mathbf{b}_3$ are the reciprocal lattice vectors, corresponding to $\mathbf{a}_1$, $\mathbf{a}_2$, $\mathbf{a}_3$, which are shown in Fig. 1(a). Huge change, up to two times, of the basic reflection (0,2,0) has been also observed in the magnetically ordered state. At 270 K, where a spontaneous moment of 0.2 $\mu_B$ along $\mathbf{a}_1$ has been measured, the Co$^{3+}$ ions in octahedral sites are considered to be in the LS state, and only the spins in pyramidal sites are ordered. The ordering is described by two wave vectors, $\mathbf{k}_{22} = \mathbf{b}_3/2$ and $\mathbf{k}_{19} = 0$. The former gives the antiferromagnetic spin component along $\mathbf{a}_2$, while the latter is for the ferromagnetic component along $\mathbf{a}_1$. The moment value is found to be $M_{Py}$ = 0.71(2) $\mu_B$, and its ferromagnetic component $M_{Px}$ = 0.50(1) $\mu_B$, which results in the spontaneous moment $M_0$ = 0.25(1) $\mu_B$/Co, since the moment of Co$^{3+}$ in pyramidal coordination is zero.

Taskin et al.[7] have concluded on the basis of magnetic measurements performed with a detwinned single crystal of GdBaCo$_2$O$_{5.50}$ that the moments in one plane with pyramidal sites are ordered ferromagnetically along [1,0,0]. The moments in the second pyramidal plane are ferromagnetically ordered in the opposite direction. Taking into account the structural data,[10] the moment in the octahedral sites in between these two planes is assumed to be zero.

This short review of controversial publications clearly demonstrates that the Co$^{3+}$ spin ordering in these materials, which one could expect to be independent on the rare earth, at least in the high-temperature range, is still unknown. Usually the data obtained from single crystals are more reliable than those from polycrystalline samples synthesized by a ceramic technology. In this case the single crystal[6] grown using very promising floating zone method and then thermally treated in the flowing oxygen to a proper oxygen content shows all superlattice reflections, which have been observed for very different oxygen contents.[12] This gives an indication that the results are very sensitive to the oxygen distribution over possible crystallographic sites and to the sample homogeneity at some average oxygen content. It seems that samples instead of the materials have been investigating up to now. We believe that, as a criterion of correct solution of the spin structure should be the same result obtained with different samples of the same material.

We present in this paper the results of neutron diffraction studies of the TbBaCo$_2$O$_{5+x}$ magnetic structure performed on two polycrystalline samples synthesized at different conditions. One sample (I) contains two ordered phases, $a_p \times 2a_p \times 2a_p$ with the oxygen content 5.49(1) and $a_p \times a_p \times 2a_p$ with statistically occupied position of the apical oxygen with $x$ = 5.39(5) as well as a phase with short-range ordering and a periodicity of 3$a_p$. The other sample (II) has a single phase $a_p \times 2a_p \times 2a_p$, and $x$ = 5.53(1). It is shown that the phase of interest, $a_p \times 2a_p \times 2a_p$, in both cases has the same spin structure. The magnetic structure is different of those reported before. Below the Néel temperature $T_N \approx$ 290 K, the spin states in the pyramidal sites are ordered ferrimagnetically, resulting in a spontaneous moment of 0.30(3) $\mu_B$ per one Co$^{3+}$ ion. Instead of spin-state ordering into the rows along [0,0,1] alternating in the [1,0,0] direction,[5] a chess-board like ordering fits our data much better. This type of spin-state ordering undoubtedly indicates that the space group that describes the unit cell $2a_p \times 2a_p \times 2a_p$ is $D_{2h}^{5}$–$Pmma$. This statement correlates with our X-ray diffraction data.[14] Very weak superstructure reflections from single crystals of GdBaCo$_2$O$_{5.50}$ and DyBaCo$_2$O$_{5.50}$ below metal-insulator transition obey an extinction law for the *Pmma* space group. It is natural to expect that the compound of Tb, which is in between Gd and Dy, should belong to the same space group. The ferrimagnetic phase exists in a narrow temperature range. It transforms into an antiferromagnetic phase at $T_1 \approx$ 255 K, with a further transition into another antiferromagnetic phase at $T_2 \approx$ 170 K. The spin states are partially ordered in all magnetic phases. Using the basis functions of irreducible representations for the symmetry groups of the wave vectors, the spin structure in each phase is determined.

Historically the sample I was the first studied, but we have solved its magnetic structure correctly only with a single-phase sample II. Having this information, the spin structure of the main phase in sample I has been found identical. Nevertheless, we present in this paper the results for both samples. Their oxygen content is slightly different, and the fact that the spin structures are the same indicates that we give the spin structure for the material rather than for a sample. Therefore we shall consider our material as TbBaCo$_2$O$_{5.5}$, indicating the actual value of $x$, when necessary.

The paper is organized as follows. In second, experimental part, first of all, the synthesis of two samples and their attestation are given. Then we present briefly the results of magnetic measurements without any discussion. At the end of this part, the instrumentation and the strategy of neutron diffraction studies are given. In third part, we present the results of the crystal structure studies. In addition to the structure of the main phase $a_p \times 2a_p \times 2a_p$, the structure of the phase $a_p \times a_p \times 2a_p$ is determined. Comparative studies of the $a_p \times 2a_p \times 2a_p$ phase at $T = 400$ K (space group $Pmmm$) and at $T = 308$ K (space group $Pmma$) are also presented. Part IV contains main results on spin structure in three magnetically ordered phases. In the process of the structure determination the symmetry analysis is used. At the end we discuss the spin ordering and its temperature transformations in terms of energy gaps between three spin states of Co$^{3+}$ and analyze possible picture of the spin-state/orbital ordering in both octahedral and pyramidal sites.

## II. EXPERIMENTAL

### A. Synthesis of the samples and their attestation

The TbBaCo$_2$O$_{5+x}$ sample I was synthesized from a mixture of high-purity oxides and carbonates Tb$_6$O$_{11}$, BaCO$_3$ and Co$_3$O$_4$ in the stoichiometric proportion. After preliminary annealing at 900°C, the pellets were thoroughly grinded in an agate mortar. The synthesis was carried out in air at 1180°C during 6 h with following cooling down to room temperature at a rate of 100 K/h. The sample has been checked using diffraction of X-rays. The neutron diffraction experiment itself has given the final answer on the sample quality. Two different perovskite-like phases of nearly the same oxygen content, which have not been observed with X-rays, are found (See Part III).

An initial sample II of TbBaCo$_2$O$_{5+x}$ was synthesized by standard solid-state reaction using Tb$_4$O$_7$, BaCO$_3$ and Co$_3$O$_4$ of a minimum purity of 99.99%. The powders were mixed and calcinated at temperatures 1000-1200°C during at least 100 h in air with several intermediate regrindings. Then, the sample was oxidized under oxygen pressure of about 60 bar at 800°C. The sample with the desired oxygen content ($x \approx 0.5$) was prepared from the oxidized sample by annealing in oxygen flow during 20 h and cooling down slowly with the rate of 1 K/min to room temperature. Oxygen content was determined by iodometric titration as 5.52(1). Averaging with the neutron diffraction result 5.54(1) at $T = 308$ K presented in Part III (Table IV), we take for the oxygen content 5.53(1). As was observed during the treatment of the neutron diffraction patterns, the sample contained some impurities that gave weak unrecognized reflections. When calculating the integrated intensities of magnetic reflections, a comparison of differential diffraction patterns with those for the sample I was useful to get rid of the spurious effects due to admixture of erroneous phases.

### B. Magnetic measurements

The DC magnetization has been measured on a Quantum Design PPMS system in the temperature range from 5 to 350 K. The magnetization field dependence for both samples [inset to Fig. 2 and curve 4 in Fig. 3(b)] shows clear evidence of the spontaneous magnetic moment in the narrow temperature range of about 250 ÷ 290 K. Second weak maximum in Fig. 3(a) indicates a phase transition with different field dependence of magnetization below this maximum (curves 5, 6) and in between two maxima (curves 3, 4).

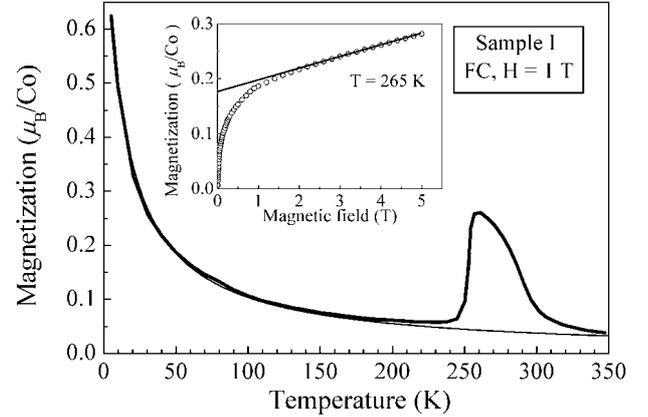

FIG. 2. Temperature dependence of magnetization measured in the field 1 T for the sample I. Thin line shows magnetization of Tb$^{3+}$. Field dependence in the maximum of magnetization is shown in the inset.

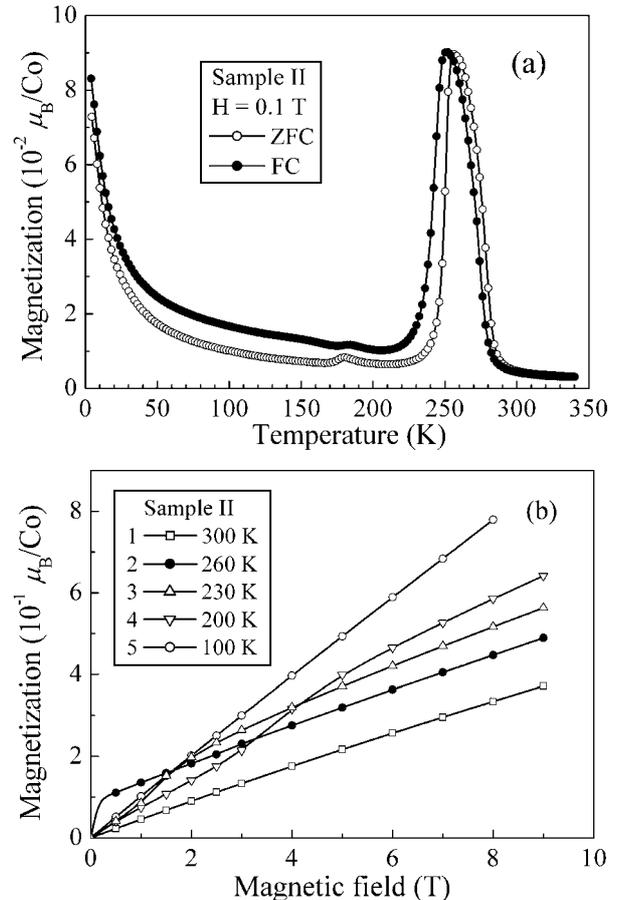

FIG. 3. (a) Temperature dependence of magnetization measured in the field 0.1 T for sample II. In addition to the main maximum at 260 K, the second weak maximum may be noticed at 180 K. (b) Isotherms measured at some characteristic temperatures (See text).

### C. Neutron diffraction experiments

Neutron diffraction experiments have been carried out on Swiss Spallation Neutron Source SINQ[15] at Paul Scherrer Institute. Two powder diffractometers have been used, HRPT[16] and DMC. The former is a powder diffractometer with the monochromator take-off angle of 120°, which provides very high resolution up to $\Delta d/d \sim 10^{-3}$. This instrument with the wave length $\lambda = 1.886$ Å or $\lambda = 1.494$ Å has been used for the crystal structure refinement. The cold neutron powder diffractometer DMC with the wave length $\lambda = 4.200$ Å is very efficient in the magnetic structure investigations.

Our strategy was the following. First, the crystal structure was refined in the paramagnetic phase by the profile analysis[17,18] on the basis of data from the HRPT diffractometer. Second, with this crystal structure the spin ordering was determined from the DMC data. Well resolved magnetic reflections on the differential diffraction patterns allowed finding the wave vector of magnetic structure. Using procedure,[19] the basis functions have been constructed for irreducible representations of the group of the wave vector. Different models of magnetic structure obtained with these basis functions were checked on the basis of integrated intensities of magnetic reflections. The data from two samples were combined, if it was desirable to improve statistics or to get rid of spurious reflections, which were different for each sample. Then the crystal structure has been refined again by means of the profile analysis of the HRPT data, with the magnetic parameters being fixed. Finally, the magnetic structure parameters have been refined once more with corrected atomic positions.

## III. CRYSTAL STRUCTURE

### A. Polymorphism due to oxygen distribution

The first attempt to fit the HRPT diffraction pattern measured in paramagnetic phase ($T = 302$ K) has failed. Commonly used structure with the unit cell $a_p \times 2a_p \times 2a_p$ described by the space group $Pmmm$ results in very poor values of the least squares residual $\chi^2 = 155$ and the profile reliability factor $R_{wp} = 8.1$. The origin of this discrepancy becomes clear from a simple intensity comparison of two reflections, (0,1,0) and (0,0,1), which are well resolved when measured on the DMC diffractometer (Fig. 4). The intensities of these reflections for the phase $a_p \times 2a_p \times 2a_p$ of $TbBaCo_2O_{5.5}$ are shown by the grey bars. The experimental intensity of (0,1,0), which is about two times weaker, indicates that there is an essential admixture of a phase with the unit cell $a_p \times a_p \times 2a_p$, since it gives no reflection in this position. A superstructure diffuse peak at the (1/3,0,0) position indicates a partially disordered phase, apparently of the type.[12] This peak is fitted by a Lorentzian with the width corresponding to the correlation length $\xi = 62(6)$ Å, as shown in the inset of Fig. 4.

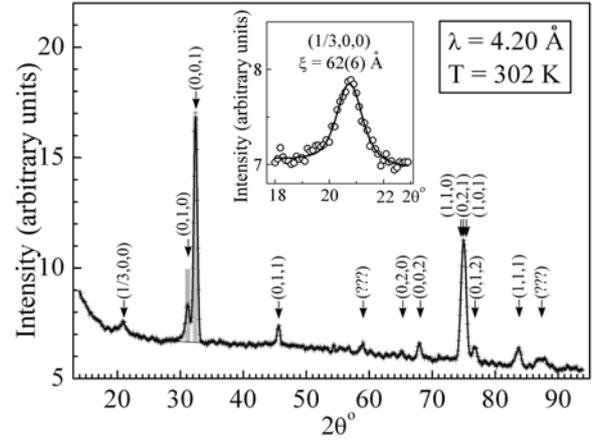

FIG. 4. Diffraction pattern from sample I measured on the DMC diffractometer at the same temperature as that in Fig. 5 below. Positions of the Bragg reflections ($h,k,l$) for the phase $a_p \times 2a_p \times 2a_p$, unrecognized reflections (???) and a superstructure peak (1/3,0,0) are indicated by arrows. A fit of the (1/3,0,0) peak by a Lorentzian is shown in the inset.

In spite of some traces of unrecognized erroneous phases [reflections marked as (???)], when using the FULLPROF program,[17,18] we have made quite successfully a two-phase profile analysis of the diffraction pattern displayed in Fig. 5.

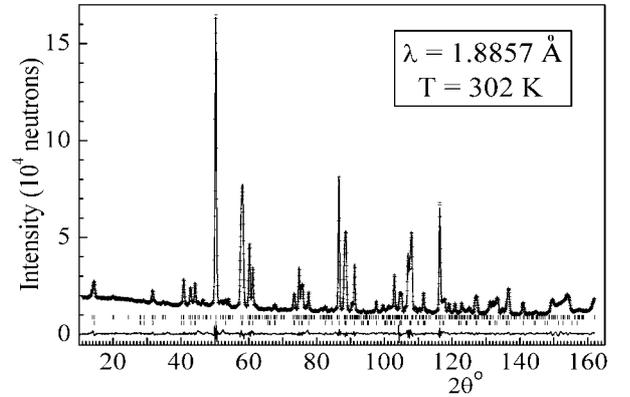

FIG. 5. Diffraction pattern from sample I above $T_N$. The best fit by two phases, $2a_p \times 2a_p \times 2a_p$ ($Pmma$) and $a_p \times a_p \times 2a_p$ ($Pmmm$).

The $\chi^2$ has been improved to $\chi^2 = 15$, which is a reasonable value for the sample of this homogeneity. For the main phase, as for the neighbor Gd and Dy materials, the space group $Pmma$ with the unit cell $2a_p \times 2a_p \times 2a_p$ has been used instead of $Pmmm$ with the unit cell $a_p \times 2a_p \times 2a_p$. Second phase has been described by the space group $Pmmm$ with the unit cell $a_p \times a_p \times 2a_p$. Corresponding to these phases, Bragg positions are indicated by the upper and the lower series of ticks below the profile. The difference between observed and calculated profiles is shown at the bottom. The phase fractions are found to be 75.7(1.1)% and 24.3(6)%, respectively. The structure parameters are collected in Table I and Table II. Occupancies of the apical oxygen sites

TABLE I. Atomic positions $x/a_1$, $y/a_2$, $z/a_3$, isotropic temperature factors $B$, and occupancies $n$ determined for the phase 1–$(a_p \times 2a_p \times 2a_p)$ TbBaCo$_2$O$_{5.49(1)}$ (sample I) at $T = 302$ K from the two-phase neutron powder profile analysis. The crystal symmetry is orthorhombic (*Pmma*; $Z = 4$), and the lattice parameters are $a_1 = 7.7404(2)$ Å, $a_2 = 7.8170(2)$ Å, $a_3 = 7.5185(2)$ Å. The phase fraction is 74(2)%.

| Atom | Position | $x/a_1$ | $y/a_2$ | $z/a_3$ | $B$ | $n$ |
|---|---|---|---|---|---|---|
| Tb | 4h | 0 | 0.2737(4) | 0.5 | 1.00(7) | 1 |
| Ba | 4g | 0 | 0.2489(8) | 0 | 1.0(1) | 1 |
| CoPy1 | 2e | 0.25 | 0 | 0.253(4) | 0.5(2) | 1 |
| CoPy2 | 2e | 0.25 | 0 | 0.735(4) | 0.5(2) | 1 |
| CoOc1 | 2f | 0.25 | 0.5 | 0.274(2) | 0.4(2) | 1 |
| CoOc2 | 2f | 0.25 | 0.5 | 0.764(3) | 0.4(2) | 1 |
| O1 | 2e | 0.25 | 0 | 0.004(3) | 0.9(2) | 0.99(3) |
| O1′ | 2e | 0.25 | 0 | 0.5 | 0 | 0.07(2) |
| O2 | 2f | 0.25 | 0.5 | 0.008(2) | 1.6(2) | 1.00(3) |
| O3 | 2f | 0.25 | 0.5 | 0.498(4) | 1.6(2) | 0.92(3) |
| O4 | 4i | −0.002(2) | 0 | 0.3108(8) | 1.55(8) | 1 |
| O5 | 4i | −0.006(2) | 0.5 | 0.2694(8) | 1.55(8) | 1 |
| O61 | 4k | 0.25 | 0.239(1) | 0.306(1) | 1.30(9) | 1 |
| O62 | 4k | 0.25 | 0.248(1) | 0.715(1) | 1.30(9) | 1 |

TABLE II. Atomic positions $x/a_1$, $y/a_2$, $z/a_3$, isotropic temperature factors $B$, and occupancies $n$ determined for the phase 2–$(a_p \times a_p \times 2a_p)$ TbBaCo$_2$O$_{5.39(5)}$ (sample I) at $T = 302$ K from the two-phase neutron powder profile analysis. The unit cell is orthorhombic (*Pmmm*; $Z = 1$), with $a_1 = 3.8928(1)$ Å, $a_2 = 3.8807(1)$ Å, $a_3 = 7.5185(2)$ Å. The phase fraction is 26(1)%.

| Atom | Position | $x/a_1$ | $y/a_2$ | $z/a_3$ | $B$ | $n$ |
|---|---|---|---|---|---|---|
| Tb | 1h | 0.5 | 0.5 | 0.5 | 0.5(1) | 1 |
| Ba | 1f | 0.5 | 0.5 | 0 | 0.5(1) | 1 |
| Co | 2q | 0 | 0 | 0.248(1) | 0.7(2) | 1 |
| O1 | 1a | 0 | 0 | 0 | 0.4(3) | 0.82(4) |
| O2 | 1c | 0 | 0 | 0.5 | 0.4(3) | 0.57(4) |
| O3 | 2r | 0 | 0.5 | 0.202(2) | 0.6(1) | 1 |
| O4 | 2s | 0.5 | 0 | 0.203(2) | 0.6(1) | 1 |

$\chi^2 = 26.5$; $R_{wp} = 3.3$; $R_{Bragg1} = 5.6$; $R_{Bragg2} = 5.0$. [Bragg1 → $(2a_p \times 2a_p \times 2a_p)$; Bragg2 → $(a_p \times a_p \times 2a_p)$]

have been varied together with the other parameters show that the oxygen content in the phase 2 is only 0.1 smaller than in the phase 1. There is no ordering of oxygen atoms and vacancies along **y** axis, which would result in the sequence of octahedra and pyramids as in the phase 1. Similar structure was reported for a fast-cooled sample of TbBaCo$_2$O$_{5.4}$.[4] The ions of Tb$^{3+}$ and Ba$^{2+}$ were found to be partially ordered, with the oxygen vacancies being predominantly located in the Tb-rich (0,0,1) planes.

### B. The crystal structure and its temperature dependence

The diffraction pattern from sample II is fitted very well by a single phase. A conventional cell $a_p \times 2a_p \times 2a_p$ and the space group *Pmmm* ($Z = 2$) have been used for $T = 400$ K. Again, taking into account our X-ray findings for the Gd and Dy materials mentioned above, we have made the profile analysis of diffraction pattern measured at $T = 308$ K on the basis of the unit cell $2a_p \times 2a_p \times 2a_p$ with the symmetry *Pmma* ($Z = 4$). The results of these refinements are collected in Table III and Table IV, respectively. Atomic coordinates are in reasonable agreement with the published results, for instance, with Ref. 9. Occupancy of the vacant position in the pyramid has been also refined, since results for the sample I indicate that a pure vacancy, apparently, never exists. Some amount of oxygen [0.14(2) at $T = 308$ K] has been found indeed in the position (1/4, 0, 1/2). [The occupancy of this position in the main phase of sample I is 0.07(2).] The other apical positions in both samples contain up to (6 ÷ 8)% of vacancies. Apparently, this leads to some difference of the lattice parameters and of the atomic positions in various samples with the same oxygen content.

TABLE III. Atomic positions $x/a_1$, $y/a_2$, $z/a_3$, isotropic temperature factors $B$, and occupancies $n$ determined for TbBaCo$_2$O$_{5.54(2)}$ (sample II) at $T = 400$ K from the neutron powder profile analysis. The crystal symmetry is orthorhombic (*Pmmm*; $Z = 2$), and the lattice parameters are $a_1 = 3.8529(1)$ Å, $a_2 = 7.8458(2)$ Å, $a_3 = 7.5563(2)$ Å. Criteria of the refinement quality are: $\chi^2 = 4.3$; $R_{wp} = 4.0$; $R_{Bragg} = 6.6$.

| Atom | Position | $x/a_1$ | $y/a_2$ | $z/a_3$ | $B$ | $N$ |
|---|---|---|---|---|---|---|
| Tb | 2p | 0.5 | 0.2678(4) | 0.5 | 0.82(4) | 1 |
| Ba | 2o | 0.5 | 0.2480(5) | 0 | 0.24(5) | 1 |
| CoPy | 2q | 0 | 0 | 0.253(1) | 0.35(9) | 1 |
| CoOc | 2r | 0 | 0.5 | 0.255(1) | 1.4(1) | 1 |
| O1 | 1a | 0 | 0 | 0 | 1.0(1) | 1.00(2) |
| O1′ | 1c | 0 | 0 | 0.5 | 1.0(1) | 0.14(2) |
| O2 | 1e | 0 | 0.5 | 0 | 1.0(1) | 1.00(2) |
| O3 | 1g | 0 | 0.5 | 0.5 | 1.0(1) | 0.94(2) |
| O4 | 2s | 0.5 | 0 | 0.3092(6) | 1.50(4) | 1 |
| O5 | 2t | 0.5 | 0.5 | 0.2694(6) | 1.50(4) | 1 |
| O6 | 4u | 0 | 0.2414(5) | 0.2960(3) | 1.29(5) | 1 |

TABLE IV. Atomic coordinates $x/a_1$, $y/a_2$, $z/a_3$, isotropic temperature factors $B$, and occupancies $n$ determined for TbBaCo$_2$O$_{5.54(1)}$ (sample II) at $T = 308$ K from the neutron powder profile analysis. The crystal symmetry is orthorhombic (*Pmma*; $Z = 4$), and the lattice parameters are $a_1 = 7.7360(2)$ Å, $a_2 = 7.8085(2)$ Å, $a_3 = 7.5331(2)$ Å. The refinement quality is given by $\chi^2 = 4.7$; $R_{wp} = 4.2$; $R_{Bragg} = 6.6$.

| Atom | Position | $x/a_1$ | $y/a_2$ | $z/a_3$ | $B$ | $n$ |
|---|---|---|---|---|---|---|
| Tb | 4h | 0 | 0.2736(4) | 0.5 | 0.45(4) | 1 |
| Ba | 4g | 0 | 0.2468(5) | 0 | −0.16(5) | 1 |
| CoPy1 | 2e | 0.25 | 0 | 0.259(4) | 0.4(1) | 1 |
| CoPy2 | 2e | 0.25 | 0 | 0.752(4) | 0.4(1) | 1 |
| CoOc1 | 2f | 0.25 | 0.5 | 0.256(5) | 1.1(1) | 1 |
| CoOc2 | 2f | 0.25 | 0.5 | 0.746(5) | 1.1(1) | 1 |
| O1 | 2e | 0.25 | 0 | 0.000(3) | 0.5(1) | 0.94(2) |
| O1′ | 2e | 0.25 | 0 | 0.5 | 0 | 0.24(2) |
| O2 | 2f | 0.25 | 0.5 | 0.002(4) | 1.9(1) | 1.00(2) |
| O3 | 2f | 0.25 | 0.5 | 0.502(4) | 1.9(1) | 0.90(2) |
| O4 | 4i | −0.003(2) | 0 | 0.3025(8) | 1.30(5) | 1 |
| O5 | 4j | 0.002(2) | 0.5 | 0.2714(8) | 1.30(5) | 1 |
| O61 | 4k | 0.25 | 0.2304(7) | 0.305(1) | 0.95(6) | 1 |
| O62 | 4k | 0.25 | 0.2523(7) | 0.710(1) | 0.95(6) | 1 |

## IV. MAGNETIC STRUCTURE

### A. Experimental data

A series of diffraction patterns measured on sample II with the wave length $\lambda = 4.200$ Å in the temperature range $10 \leq T\,(\text{K}) \leq 300$ is shown in Fig. 6. Figure 7 displays intensity temperature dependence of three magnetic reflections that characterize best of all the temperature evolution of magnetic structure. One can definitely see three phase transitions. Reflection (1,1,1) indicates that magnetic ordering appears at the Néel temperature $T_N \approx 290$ K, and the magnetic unit cell coincides with the crystallographic one $2a_p \times 2a_p \times 2a_p$ (*Pmma*). This means that the translational symmetry of magnetic structure is given by the wave vector $\mathbf{k}_{19} = 0$. The peak intensity $I_{(111)}(T)$ follows very well the temperature dependence of magnetization shown in Fig. 3(a).

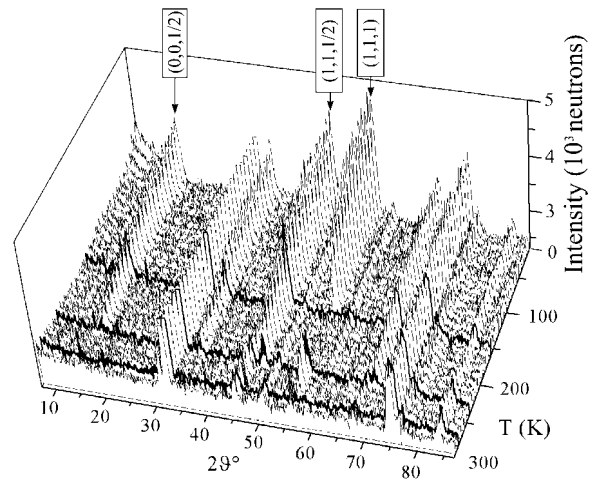

FIG. 6. Diffraction patterns from sample II collected on DMC with a temperature step of 10 K.

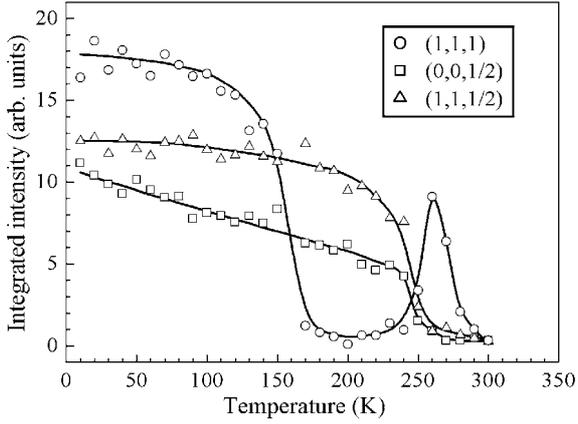

FIG. 7. Intensity temperature dependence of reflections (1,1,1), (0,0,1/2), and (1,1,1/2). Indices are given for the unit cell $2a_p \times 2a_p \times 2a_p$ (*Pmma*).

At $T_1 \approx 255$ K, intensity of the (1,1,1) reflection abruptly decreases to zero, and simultaneously reflections (0,0,1/2), (0,1,1/2) arise. This indicates antiferromagnetic doubling of the magnetic unit cell along $\mathbf{a}_3$ axis. The wave vector of this antiferromagnetic structure is $\mathbf{k}_{22} = \mathbf{b}_3/2$. Data for the magnetic and crystal structure refinement in this phase have been obtained at 230 K. A crystallographic unit cell $2a_p \times 2a_p \times 2a_p$ (*Pmma*) is confirmed.

A second transition occurs at $T_2 \approx 170$ K, close to the weak peak of magnetization at 178 K [Fig. 3(a)]. Antiferromagnetic reflections (0,0,1/2) and (0,1,1/2) exhibit no visible change at this temperature, while reflection (1,1,1) appears again. Former two reflections exist if a spin direction reverses when being translated by $\mathbf{a}_3$, while the latter vanishes at this antitranslation. This discrepancy is removed, if the unit cell edge $a_3$ of the *Pmma* phase is doubled at $T_2$ due to a structural phase transition with the wave vector $\mathbf{k}_{22} = \mathbf{b}_3/2$. Then the indices of reflections (0,0,1/2), (0,1,1/2) and (1,1,1) are transformed into (0,0,1), (0,1,1) and (1,1,2), respectively. From now, when indexing reflections and calculating their intensity, we always consider the big unit cell $2a_p \times 2a_p \times 4a_p$. The wave vector of magnetic structure for this unit cell in the low-temperature phase is $\mathbf{k}_{19} = 0$. According to Ref. 20, the highest sub-group that fits a structural transition with $\mathbf{k}_{22}$ and with small distortions of the *Pmma* phase is $D_{2h}^8 - Pcca$. The structure parameters obtained by the profile refinement at $T = 100$ K are collected in Table V.

Since the high-temperature phase is of the main interest, we pay more attention to the experimental data obtained at the magnetization maximum 260 – 265 K. The differential patterns with those measured above $T_N$ are presented in Fig. 8. When comparing the patterns (a) and (b) for the samples I and II, respectively, one can see a series of additional reflections on the first pattern, apparently, due to the phase $a_p \times a_p \times 2a_p$. Nevertheless, common for both samples magnetic reflections, when normalized, have equal intensities in the limits of a standard deviation. Therefore, we have combined them to obtain better statistic. In particular, very important are reflections (0,1,0) and (0,1,1) shown in Figs. 8(c) and 8(d), respectively. They indicate that the ferromagnetic components arise in the part $a_p \times 2a_p \times 2a_p$ of the unit cell.

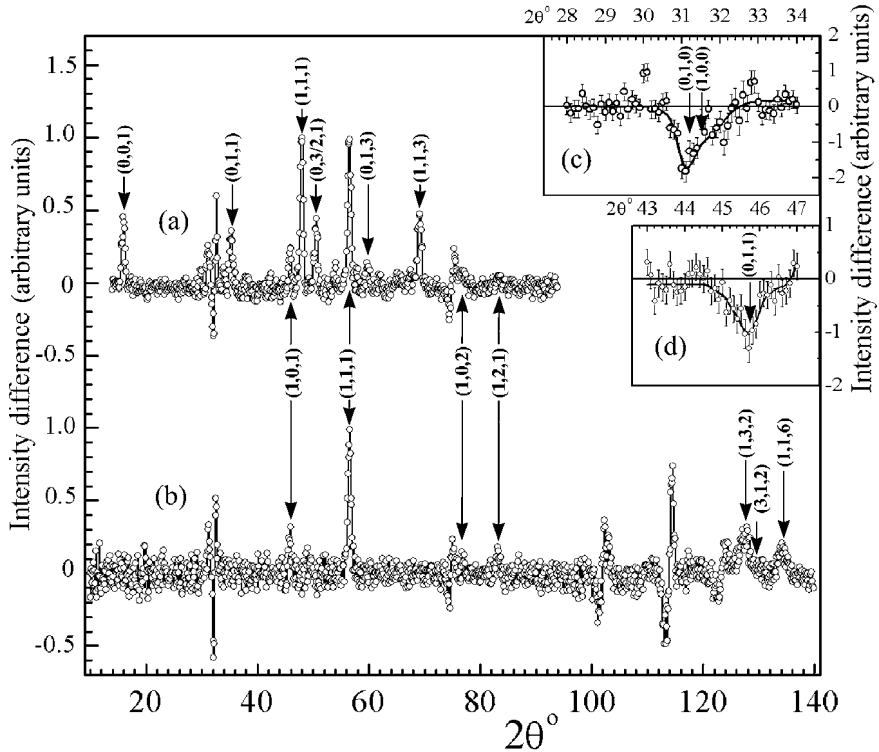

FIG. 8. Differential diffraction patterns: (a) $I(265$ K$) - I(300$ K$)$ sample I; (b) $I(260$ K$) - I(308$ K$)$ sample II; (c), (d) $I(255$ K$) - I(265$ K$)$ sample I. Miller indices are given for the unit cell $2a_p \times 2a_p \times 4a_p$.

TABLE V. Atomic coordinates $x/a_1$, $y/a_2$, $z/a_3$, isotropic temperature factors $B$, and occupancies $n$ of crystallographic positions determined for TbBaCo$_2$O$_{5.53}$ (sample II) at $T = 100$ K from the neutron powder profile analysis. Orthorhombic space group is $Pcca$ ($Z = 8$), and the lattice parameters are $a_1 = 7.7359(2)$ Å, $a_2 = 7.7900(2)$ Å, $a_3 = 15.0263(5)$ Å. The refinement quality is characterized by $\chi^2 = 5.6$; $R_{wp} = 4.1$; $R_{Bragg} = 6.7$.

| Atom | Position | $x/a_1$ | $y/a_2$ | $z/a_3$ | $B$ | $n$ |
|---|---|---|---|---|---|---|
| Tb1 | 4c | 0 | 0.271(1) | 0.25 | 0.70(5) | 1 |
| Tb2 | 4c | 0 | 0.723(1) | 0.25 | 0.70(5) | 1 |
| Ba | 8f | 0.005(3) | 0.2502(6) | −0.0032(6) | −0.33(8) | 1 |
| CoPy1 | 4d | 0.25 | 0 | 0.135(1) | 0.0(1) | 1 |
| CoPy2 | 4d | 0.25 | 0 | 0.376(2) | 0.0(1) | 1 |
| CoOc1 | 4e | 0.25 | 0.5 | 0.125(2) | 1.4(2) | 1 |
| CoOc2 | 4e | 0.25 | 0.5 | 0.376(3) | 1.4(2) | 1 |
| O1 | 4d | 0.25 | 0 | 0.000(1) | 0.16(8) | 0.94 |
| O1′ | 4d | 0.25 | 0 | 0.2500 | 0 | 0.24 |
| O2 | 4e | 0.25 | 0.5 | 0.001(2) | 1.9(1) | 1.00 |
| O3 | 4e | 0.25 | 0.5 | 0.252(2) | 1.9(1) | 0.90 |
| O4 | 8f | −0.003(2) | 0.009(2) | 0.1498(4) | 1.06(7) | 1 |
| O5 | 8f | 0.004(2) | 0.486(1) | 0.1338(4) | 1.06(7) | 1 |
| O61 | 8f | 0.256(3) | 0.2319(8) | 0.1552(5) | 0.74(7) | 1 |
| O62 | 8f | 0.252(2) | 0.2538(8) | 0.8555(5) | 0.74(7) | 1 |

### B. Symmetry analysis

Knowing the space group $G$ and the wave vector $\mathbf{k}$ of magnetic structure in all magnetically ordered phases, one can find the basis functions $\mathbf{S}_{i\upsilon}^{\lambda\mu}$ of irreducible representations of the group $G_\mathbf{k}$, where $i$, $\nu$ and $\lambda\mu$ numerate atom, representation and element of the representation matrix, respectively. This procedure is described in detail in Ref. 19. Three magnetically ordered phases are numerated as 1 - (255 K < $T$ < 290 K), 2 - (170 K < $T$ < 255 K), and 3 - ($T$ < 170 K).

#### 1. Phase 1

Space group $D_{2h}^5$ – $Pmma$. Four Co$^{3+}$ ions in pyramidal coordination occupy two-fold sites (Py1, Py2) 2$e$: 1–(1/4, 0, $z$); 2–(3/4, 0, –$z$). Four Co$^{3+}$ ions in octahedral coordination occupy two-fold sites (Oc1, Oc2) 2$f$: 1–(1/4, 1/2, $z$); 2–(3/4, 1/2, –$z$). The coordinate $z$ for every pair is indicated in Table IV as determined at $T = 308$ K. The wave vector of magnetic structure is found to be $\mathbf{k}_{19} = 0$. The small group of this wave vector is

$$G_\mathbf{k} \in g_1, g_2, g_3, g_4, g_{25}, g_{26}, g_{27}, g_{28}, \quad (1)$$

where a symmetry element $g$ is denoted according to Wigner-Seitz as $\{h|\tau_h\}$, with $h$ and $\tau_h$ being a rotational element at the unit cell origin and an accompanying translation, respectively. The numbering of $g$ corresponds to Ref. 13, where one can find the matrices of $h$. It should be pointed out that rotational part of $G_\mathbf{k}$ is the same for all three magnetically ordered phases. Magnetic representation for every pair of equivalent Co ions, which is a matrix $D_M^{\mathbf{k}_{19}}$ of dimensionality $6 \times 6$, can be reduced by means of Eqs. (8.21) and (8.22) of Ref. 19 to 1D ($\lambda = \mu = 1$) irreducible representations $\tau_\nu$ given in Table 32 of the handbook 13 as

$$D_M^{\mathbf{k}_{19}} = \tau_2 + \tau_3 + \tau_4 + \tau_5 + \tau_7 + \tau_8. \quad (2)$$

Expression of $D_M^{\mathbf{k}_{19}}$ through $\tau_\nu$ is the same for both pyramidal and octahedral sublattices as well as the basis functions, which are obtained by means of Eq. (9.15)[19] and collected in Table VI as projections of the unit vectors $\mathbf{S}_i$ on the crystal axes.

TABLE VI. Basis functions of irreducible representations for the pyramidal and octahedral sub-lattices of the phase 1 – $Pmma$.

| $\nu$ | 2 | | 3 | | 4 | | 5 | | 7 | | 8 | |
|---|---|---|---|---|---|---|---|---|---|---|---|---|
| $i$ | 1 | 2 | 1 | 2 | 1 | 2 | 1 | 2 | 1 | 2 | 1 | 2 |
| $S_x$ | 0 | 0 | 0 | 0 | 1 | $\bar{1}$ | 0 | 0 | 1 | 1 | 0 | 0 |
| $S_y$ | 0 | 0 | 1 | 1 | 0 | 0 | 0 | 0 | 0 | 0 | 1 | $\bar{1}$ |
| $S_z$ | 1 | $\bar{1}$ | 0 | 0 | 0 | 0 | 1 | 1 | 0 | 0 | 0 | 0 |

A very important conclusion, which follows from this analysis, concerns the nature of spontaneous moment. Weak ferromagnetism due to Dzyaloshinskii-Moriya spin canting, as has been suggested in a number of publications,[21-23] is allowed if a one-dimensional irreducible representation enters $D_M^\mathbf{k}$ at least twice, with one basis function being composed of a ferromagnetically arranged couple of unit vectors $\mathbf{S}_{i\nu}$. In our case a spontaneous moment can appear only from different spin values either in non-equivalent couples in one sublattice, or in different sublattices. In other words, there should be a ferrimagnetic ordering.

#### 2. Phase 2

This phase appears due to a change of the wave vector to $\mathbf{k}_{22} = \mathbf{b}_3/2 = (2\pi/a_3)[0,0,1/2]$ in the frame of the same space group $Pmma$. This means that direction of each spin is reversed when being translated by $\mathbf{a}_3$. Therefore, the mutual spin orientation in a half of magnetic unit cell has to be analyzed. The magnetic representation of the same dimensionality as in the previous case is reduced to 2D

irreducible representations obtained from Table 53 of the handbook 13 by means of Eq. (2.32)[19] as

$$D_M^{\mathbf{k}_{22}} = \tau_1 + 2\tau_2. \qquad (3)$$

Though the matrices of these representations are complex, they can be made real by a unitary transform using matrices

$$U = \frac{1}{2}\begin{pmatrix} 1+i & -1+i \\ 1-i & -1-i \end{pmatrix}; \quad U^{-1} = \frac{1}{2}\begin{pmatrix} 1-i & 1+i \\ -1-i & -1+i \end{pmatrix}. \qquad (4)$$

Being the same for the pyramidal and the octahedral sublattices, the basis functions corresponding to every 2D irreducible representation and similar to those from Table VI are shown in Table VII.

TABLE VII. Basis functions $\mathbf{S}_{i\upsilon}^{\lambda\mu}$ of irreducible representations for the pyramidal and octahedral sublattices of the phase 2. Space group $Pmma$, wave vector $\mathbf{k}_{22} = \mathbf{b}_3/2$. The indices $\lambda\nu$ and $i$ numerate element of a 2D matrix and an atom, respectively.

| | | \multicolumn{8}{c}{$\mathbf{S}_{i\upsilon}^{\lambda\mu}$} |
|---|---|---|---|---|---|---|---|---|---|
| $\nu$ | $\lambda,\mu\to$ | 1,1 | | 1,2 | | 2,1 | | 2,2 | |
| $\downarrow$ | $i\to$ | 1 | 2 | 1 | 2 | 1 | 2 | 1 | 2 |
| 1 | $S_x$ | 0 | 0 | 0 | 0 | | | | |
| | $S_y$ | 1 | 1 | 1 | $\bar{1}$ | | | | |
| | $S_z$ | 0 | 0 | 0 | 0 | | | | |
| 2 | $S_x$ | 1 | 1 | 0 | 0 | 1 | $\bar{1}$ | 0 | 0 |
| | $S_y$ | 0 | 0 | 0 | 0 | 0 | 0 | 0 | 0 |
| | $S_z$ | 0 | 0 | 1 | 1 | 0 | 0 | 1 | $\bar{1}$ |

Six basis functions, corresponding to different $\lambda,\nu$ of matrices for $\tau_1$ and $\tau_2$, allow a variety of magnetic structures. A single basis function describes a collinear spin arrangement of two spins, parallel or antiparallel, along one of the crystal axes. Mixing of two and more functions may result in a Dzyaloshinskii-Moriya spin canting, ferromagnetic or antiferromagnetic. (However, one should remember that this phase is antiferromagnetic because of antitranslation $\mathbf{a}_3$.)

*3. Phase 3*

Space group $D_{2h}^5 - Pcca$. The crystallographic unit cell is doubled along $\mathbf{z}$. Eight $Co^{3+}$ ions in pyramidal coordination occupy two non-equivalent four-fold sites (Py1 and Py2) $4d$: 1–(1/4, 0, $z$); 2–(3/4, 0, $-z$); 3–(1/4, 0, 1/2+$z$); 4–(3/4, 0, 1/2–$z$). Eight $Co^{3+}$ ions in octahedral coordination occupy two non-equivalent four-fold sites (Oc1 and Oc2) $4e$: 1–(1/4, 1/2, $z$); 2–(3/4, 1/2, $-z$); 3–(1/4, 1/2, 1/2+$z$); 4–(3/4, 1/2, 1/2–$z$). The coordinate $z$ for every four equivalent atoms is given in Table VI as determined at $T = 100$ K. The correspondence between these numbers and those indicated in Fig. 9 below for all magnetic phases is given in Table VIII, where the symbols Py1, Py2 and Oc1, Oc2 indicate non-equivalent pyramidal and octahedral sites, respectively. The wave vector of magnetic structure is $\mathbf{k}_{19} = 0$. Eight 1D irreducible representations $\tau_\nu$ coincide with those for the phase 1, and enter $m$ times the magnetic representation of the dimensionality $12 \times 12$ as

$$D_M^{\mathbf{k}_{19}} = \tau_1 + 2\tau_2 + \tau_3 + 2\tau_4 + 2\tau_5 + \tau_6 + 2\tau_7 + \tau_8. \qquad (5)$$

Corresponding basis functions are collected in Table VIII.

TABLE VIII. Basis functions $\mathbf{S}_{i\upsilon}$ of irreducible representations for the pyramidal and octahedral sublattices of the phase 3. Space group $Pcca$, wave vector $\mathbf{k}_{19} = 0$.

| $\nu$ | $m\to$ | \multicolumn{4}{c}{1} | \multicolumn{4}{c}{2} |
|---|---|---|---|---|---|---|---|---|---|
| $\downarrow$ | $i\to$ | 1 | 2 | 3 | 4 | 1 | 2 | 3 | 4 |
| | Py1 | 1 | 12 | 9 | 4 | 1 | 12 | 9 | 4 |
| | Py2 | 2 | 11 | 10 | 3 | 2 | 11 | 10 | 3 |
| | Oc1 | 5 | 16 | 13 | 8 | 5 | 16 | 13 | 8 |
| | Oc2 | 6 | 15 | 14 | 7 | 6 | 15 | 14 | 7 |
| 1 | $S_x$ | 0 | 0 | 0 | 0 | | | | |
| | $S_y$ | 1 | 1 | $\bar{1}$ | $\bar{1}$ | | | | |
| | $S_z$ | 0 | 0 | 0 | 0 | | | | |
| 2 | $S_x$ | 0 | 0 | 0 | 0 | 1 | $\bar{1}$ | $\bar{1}$ | 1 |
| | $S_y$ | 0 | 0 | 0 | 0 | 0 | 0 | 0 | 0 |
| | $S_z$ | 1 | $\bar{1}$ | 1 | $\bar{1}$ | 0 | 0 | 0 | 0 |
| 3 | $S_x$ | 0 | 0 | 0 | 0 | | | | |
| | $S_y$ | 1 | 1 | 1 | 1 | | | | |
| | $S_z$ | 0 | 0 | 0 | 0 | | | | |
| 4 | $S_x$ | 0 | 0 | 0 | 0 | 1 | $\bar{1}$ | 1 | $\bar{1}$ |
| | $S_y$ | 0 | 0 | 0 | 0 | 0 | 0 | 0 | 0 |
| | $S_z$ | 1 | $\bar{1}$ | $\bar{1}$ | 1 | 0 | 0 | 0 | 0 |
| 5 | $S_x$ | 0 | 0 | 0 | 0 | 1 | 1 | $\bar{1}$ | $\bar{1}$ |
| | $S_y$ | 0 | 0 | 0 | 0 | 0 | 0 | 0 | 0 |
| | $S_z$ | 1 | 1 | 1 | 1 | 0 | 0 | 0 | 0 |
| 6 | $S_x$ | 0 | 0 | 0 | 0 | | | | |
| | $S_y$ | 1 | $\bar{1}$ | $\bar{1}$ | 1 | | | | |
| | $S_z$ | 0 | 0 | 0 | 0 | | | | |
| 7 | $S_x$ | 0 | 0 | 0 | 0 | 1 | 1 | 1 | 1 |
| | $S_y$ | 0 | 0 | 0 | 0 | 0 | 0 | 0 | 0 |
| | $S_z$ | 1 | 1 | $\bar{1}$ | $\bar{1}$ | 0 | 0 | 0 | 0 |
| 8 | $S_x$ | 0 | 0 | 0 | 0 | | | | |
| | $S_y$ | 1 | $\bar{1}$ | 1 | $\bar{1}$ | | | | |
| | $S_z$ | 0 | 0 | 0 | 0 | | | | |

C. Determination of magnetic structure

The basis functions obtained in the previous section have been used to construct the models of magnetic structure for comparison of calculated and experimental intensities. In all phases, the intensity for the unit cell doubled along $\mathbf{z}$ axis has been calculated as[19]

$$I(\mathbf{Q}) \propto |\mathbf{F}(\mathbf{Q})|^2 - |\mathbf{eF}(\mathbf{Q})|^2 \qquad (6)$$

and averaged over the orthorhombic twins. Here $\mathbf{F}(\mathbf{Q})$ is the magnetic structure amplitude, $\mathbf{Q}$ is the neutron transferred momentum, and $\mathbf{e} = \mathbf{Q}/Q$. The integrated intensities from the differential patterns measured on the DMC diffractometer have been used for the experimental values. Spin structures for all phases are described in Fig. 9.

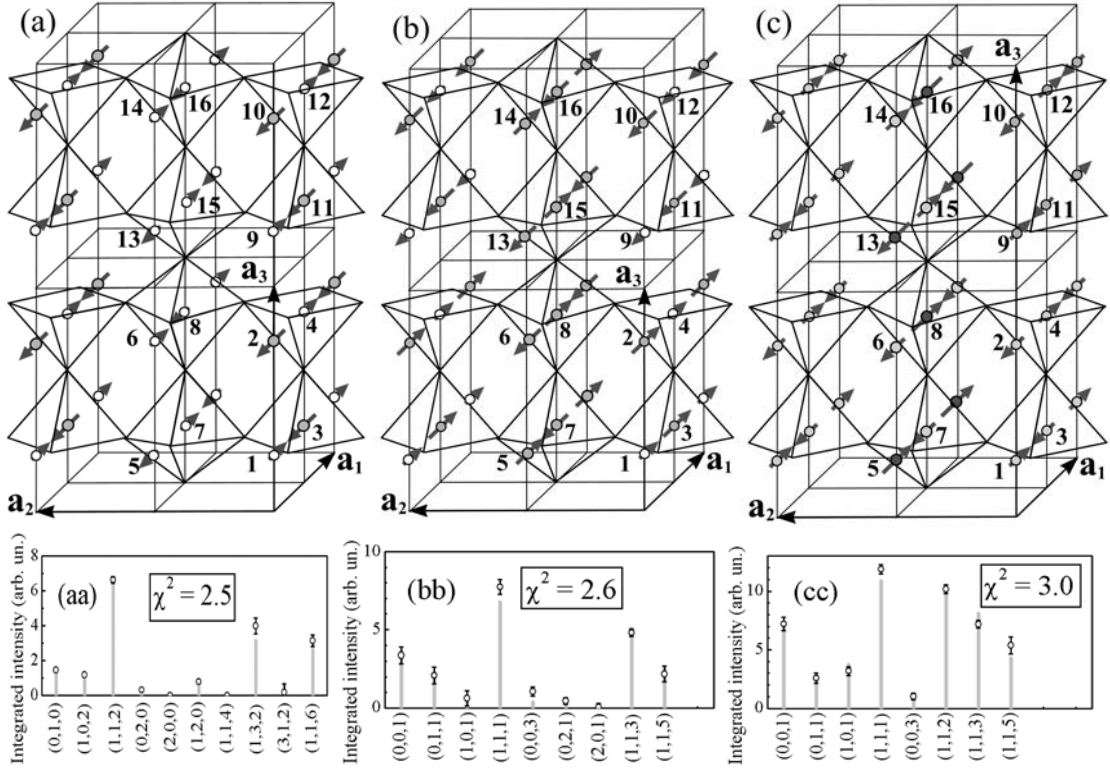

FIG. 9. Ordering of the $Co^{3+}$ moments $\mathbf{M}_n$ in three phases of $TbBaCo_2O_{5.53}$: (a) ferrimagnetic phase 1 at $T = 260$ K; (b) antiferromagnetic phase 2 at $T = 230$ K; (c) antiferromagnetic phase 3 at $T = 230$ K. The circle shadowing and the arrow length reflect the different moment value. The vectors $\mathbf{a}_1$, $\mathbf{a}_2$, $\mathbf{a}_3$ define the crystallographic unit cell for each phase. Experimental and calculated intensities (aa), (bb), (cc) correspond to the structures (a), (b), (c).

In all phases, the spins are collinear along $\mathbf{x}$ axis. The experimental (open circles) and the calculated (grey bars) intensities corresponding to the best fit are shown in Figs. 9(aa), 9(bb), and 9(cc) below each structure model. A number of other reasonable models composed of the basis functions have been tested for each phase, but with the residual squares $\chi^2 > 10$, which means that these models are improbable, in comparison with those described in Fig. 9. The models of spin ordering suggested before have been also tested with the following results, which unambiguously evidence that these models do not fit the crystal structure shown in Fig. 1(a):

- F. Fauth et al.[5] – $\chi^2 = 20$.
- M. Soda et al.[6] – $\chi^2 = 89$.
- A.A. Taskin et al.[7] – No reflections arising due to suggested unit cell doubling along $\mathbf{y}$ axis have been observed.

The moment values obtained as a result of the best fit are collected in Table IX for all 16 $Co^{3+}$ ions under the numbers $N = 1 \div 4$, $9 \div 12$ and $N = 5 \div 8$, $13 \div 16$ for the Py1, Py2 and Oc1, Oc2 sites, respectively, in accordance with Fig. 9. Except phase 3, both sublattices are composed of the same basis functions. For the phase 1 (260 K), this is a single basis function of irreducible representation $\tau_7$ (Table VI). Phase 2 (230 K) is described by the first pair of $S_x$, $S_y$, $S_z$ for irreducible representation with $\nu = 2$, as shown in Table VII.

TABLE IX. Moment values $M_n$ for the magnetic structures shown in Fig. 9. The columns (a), (b), (c) correspond to Figs. 9(a), 9(b), 9(c), respectively. For the phases 1 (260 K) and 2 (230 K) $z$-coordinates should be divided by 2.

| N | x, y, z | $M_n$ ($\mu_B$) | | |
|---|---|---|---|---|
| | | (a) 260 K | (b) 230 K | (c) 100 K |
| 1 | 1/4 0 $z_{Py1}$ | 0.46(9) | 0.28(8) | 1.76(6) |
| 2 | 1/4 0 $z_{Py2}$ | −1.65(9) | 1.22(8) | −1.76 |
| 3 | 3/4 0 1/2−$z_{Py2}$ | −1.65 | 1.22 | −1.76 |
| 4 | 3/4 0 1/2−$z_{Py1}$ | 0.46 | 0.28 | 1.76 |
| 5 | 1/4 1/2 $z_{Oc1}$ | −0.36(9) | 1.39(8) | 3.41(8) |
| 6 | 1/4 1/2 $z_{Oc2}$ | 0.39(9) | −1.52(8) | −1.47(8) |
| 7 | 3/4 1/2 1/2−$z_{Oc2}$ | 0.39 | −1.52 | −1.47 |
| 8 | 3/4 1/2 1/2−$z_{Oc1}$ | −0.36 | 1.39 | 3.41 |
| 9 | 1/4 0 1/2+$z_{Py1}$ | 0.46 | −0.28 | 1.76 |
| 10 | 1/4 0 1/2+$z_{Py2}$ | −1.65 | −1.22 | −1.76 |
| 11 | 3/4 0 −$z_{Py2}$ | −1.65 | −1.22 | −1.76 |
| 12 | 3/4 0 −$z_{Py1}$ | 0.46 | −0.28 | 1.76 |
| 13 | 1/4 1/2 1/2+$z_{Oc1}$ | −0.36 | −1.39 | −3.41 |
| 14 | 1/4 1/2 1/2+$z_{Oc2}$ | 0.39 | 1.52 | 1.47 |
| 15 | 3/4 1/2 −$z_{Oc2}$ | 0.39 | 1.52 | 1.47 |
| 16 | 3/4 1/2 −$z_{Oc1}$ | −0.36 | −1.39 | −3.41 |

As seen in Table VIII, magnetic structure of the pyramidal sublattice in the phase 3 belongs to irreducible representation with $\nu = 7$, $m = 2$, while for the octahedral one $\nu = 2$, $m = 2$. We should point out that the moments in the limits of one sublattice, even for different sublattices, are transformed according to a single irreducible representation. The question of a single irreducible representation has been discussed in Ref. 19.

## V. DISCUSSION

First objective of this work was to study the spin ordering independently of a sample. Two samples prepared by different technologies exhibit magnetic diffraction patterns, which can be interpreted using completely different magnetic structures. It is shown that sample I consists of two phases with the same content of oxygen, but not the same degree of its ordering among the apical positions. This leads to different crystal structures described in Fig. 1. Magnetic diffraction pattern from this sample is nothing more, but a sum of two patterns due to magnetic ordering in two crystallographic phases. As the unit cell parameters have always a common multiple, the unit cell parameter $a_p$ of the pseudocubic perovskite cell, one can easily take additional magnetic reflections for those of the main phase. A wrong magnetic structure can be suggested as a result. This can even happen when studying a single crystal, as far as the phase separation depends on the degree of oxygen ordering.

In our case, sample II that has been annealed in the oxygen flow has a single double perovskite phase with the unit cell at 400 K $a_p \times 2a_p \times 2a_p$ [Fig. 1(a)] with only a small part of vacant sites in pyramids occupied by oxygen. (Some unrecognized weak reflections have been observed, but they do not change drastically the differential diffraction patterns.) The perfection of oxygen ordering, although not ideal, develops itself in both magnetic and neutron diffraction measurements by three phase transitions at about 290 K, 255 K, and 170 K. Its differential diffraction pattern $I(265\ K) - I(300\ K)$ contains only one type of magnetic reflections corresponding to the $a_1 \approx 2a_p$, with the cell doubling being a result of the structural phase transition from $Pmmm$ to $Pmma$ symmetry above the temperature of magnetic ordering at $T_N \approx 290$ K. The quality of the profile refinement for $Pmmm$ and $Pmma$ is essentially the same for $T = 308$ K, as seen in Tables III and IV. The main argument for the symmetry change follows from the magnetic structure in the high-temperature magnetic phase ($T_1 < T < T_N$). We take into account also the results of Ref. 14, where this transition has been directly observed by appearance of the superstructure X-ray reflections in GdBaCo$_2$O$_{5.5}$ and DyBaCo$_2$O$_{5.5}$ (Remind that Tb is in between Gd and Dy in the periodic table). With this crystallographic unit cell the wave vector of magnetic structure in the high-temperature magnetic phase is $\mathbf{k}_{19} = 0$. Consequently, all additional reflections observed for the sample II at 265 K and shown in Fig. 8(a) are due to magnetic ordering in the erroneous phase $a_p \times a_p \times 2a_p$ [Fig. 1(b)]. Actually, the quality of the main phase $2a_p \times 2a_p \times 2a_p$ ($Pmma$) in the sample I is as good as in the sample II. Although there is no second maximum at about 170 K on the magnetization temperature dependence [Fig. 2(a)], the intensity behavior of the magnetic reflection (1,1,1) is similar to that shown in Fig. 7 for sample II. We should remind that in Ref. 5 this reflection, which was attributed to the spin-state ordering, appeared at $T_{SSO} \approx 150$ K and gradually fell with the temperature decrease. The reflections with $l = n+1$, as indexed in the doubled cell, appeared at $T_N \approx 275$ K, which was naturally considered as the point of antiferromagnetic ordering, with the spontaneous moment being ignored.

As follows from Table IX, the spontaneous moment is a result of ferrimagnetism, i.e., antiparallel spin ordering in two non-equivalent pyramidal positions. (Small moment can also result from the octahedral sites.) As discussed in the section IV-B-1, any ferromagnetic spin canting of the Dzyaloshinskii-Moriya type is forbidden by symmetry. The moment value that is obtained from the neutron diffraction data at $T = 260 \div 265$ K is $M_P = 0.60(6)$ $\mu_B$ per one Co$^{3+}$ ion in the pyramidal sublattice, or $M_0 = 0.30(3)$ $\mu_B$ per a Co$^{3+}$ ion in general. The moment is parallel to the [1,0,0] direction. Unfortunately, we cannot compare this value either with $M \approx 0.18$ $\mu_B$ (Fig. 2) or $M \approx 0.08$ $\mu_B$ [Fig. 3(a)] measured on the polycrystalline samples I and II, respectively. Apparently, a proper value is $M_0 \approx 0.35$ $\mu_B$ that has been obtained[7] by extrapolation to zero field of magnetization measured along the [1,0,0] axis on a detwinned single crystal of GdBaCo$_2$O$_{5.5}$.

We suppose that metamagnetic character of the isotherms in Fig. 3(b) is a consequence of some features of antiferromagnetic structure in the phases 2 and 3. As seen in Table IX, the absolute moment values of all the atoms in pyramidal sublattice are equal in the phase 3. However, they are distributed over two symmetrically non-equivalent sites, $N_{Py1} = 1, 4, 9, 12$ and $N_{Py2} = 2, 3, 10, 11$, which means that this is an accidental degeneration, i.e., the moments of two groups may be slightly different. Taking into account that non-equivalent spins are ordered antiferromagnetically, a net moment can appear in magnetic field of a few Tesla. Similar situation exists for the pairs of atoms $N_{Oc1} = 5, 8$ (13, 16) and $N_{Oc2} = 6, 7$ (14, 15) in octahedral sublattice of the phase 2. The field of metamagnetic transition in the phase 2 varies from ~1 T at $T = 230$ K to ~3 T at $T = 200$ K, with the moment of about 1.3 $\mu_B$/Co when extrapolated to zero field. A metamagnetic transition can be also envisaged in the field of about 8 T in the phase 3 at 150 K, but one needs higher magnetic field to estimate the extrapolated moment. Unlike phase 1, some irreducible representations enter the magnetic one twice in the phases 2 [Eq. (3)] and 3 [Eq. (5)]. This allows a canting of the spins, ferromagnetic or antiferromagnetic (hidden). For instance, the main spin component of phase 3 ordered according to $\tau_{72}$ ($S_x$, $S_x$, $S_x$, $S_x$) in the pyramidal sites and $\tau_{42}$ ($S_x$, $-S_x$, $S_x$, $-S_x$) in the octahedral sites may be accompanied by $\tau_{71}$ ($S_z$, $S_z$, $-S_z$, $-S_z$) and $\tau_{42}$ ($S_z$, $S_z$, $S_z$, $S_z$), respectively. These additional components are due to relativistic interactions, and $S_z/S_x$ should be of the order of $\Delta g/g$, where $g = 2$ for the purely spin value of the moment. In this case a weak ferromagnetism of Dzyaloshinskii-Moriya is allowed along [0,0,1] due to the octahedral spin canting.

We assume that both crystallographic phases with superstructures, either $2a_p \times 2a_p \times 2a_p$ (*Pmma*) or $2a_p \times 2a_p \times 4a_p$ (*Pcca*) appear due to ordering of spin/orbital states of the $Co^{3+}$ ions as predicted by Khomskii (Ref. 24 and Refs. therein). Radaelli and Cheong[25] have argued that difference of ionic radii due to the $Co^{3+}$ electronic structure should give too weak effects to be observed in a diffraction experiment. This is certainly the case for spherically symmetric ions. Ordering of orbitals with different shape may result in pronounced difference of interatomic distances for the non-equivalent sites, as has been shown in our previous work.[14] The distances from the non-equivalent cobalt ions in pyramidal (Py1, Py2) and octahedral (Oc1, Oc) sites to their ligands are collected in Table X together with the moment values. The moment should increase with the temperature due to excitation of the intermediate and the high spin states. On the other hand, as usually, the moment value should fall when nearing the Néel temperature. This is a reason why it is not easy to assign the spin state to an atom from its moment value. Nevertheless, the situation at $T = 100$ K is very clear. It is obvious that $Co^{3+}$ ion in the Oc1 site is in the HS state ($t_{2g}^4 e_g^2$, $S = 2$), while all the rest have the IS state ($t_{2g}^5 e_g^1$, $S = 1$). At $T = 230$ K, the Py1 $Co^{3+}$ ion should be in the LS state ($t_{2g}^6 e_g^0$, $S = 0$), while all the other ions are most probably in the IS state. Taking into account that at $T = 260$ K the moments should be strongly decreased due to the $T_N$ vicinity, we can hardly assign a definite spin state. Apparently, the Py2 $Co^{3+}$ ion is in the HS state. The other pyramidal and octahedral ions should be in the ground singlet state with an admixture either of the IS or HS state. The 260 K results are in contradiction with the conclusion[10] made for the Gd material on the IS state in pyramidal sublattice and the LS state in the octahedral one. The spin states as well as the spin structure disagree with the conclusions[7] that has been also made for the Gd material. (We suppose that the rare earth influence on the $Co^{3+}$ spin and spin-state ordering is negligible at this temperature.)

TABLE X. The distances (Å) from the $Co^{3+}$ ions in the non-equivalent sites, pyramidal (Py1, Py2) and octahedral (Oc1, Oc2), to the ligands in comparison with the magnetic moment values $M$ ($\mu_B$) at $T = 260$ K (phase 1, *Pmma*), $T = 230$ K (phase 2, *Pmma*), and $T = 100$ K (phase 3, *Pcca*)

| $T$(K) | 260 | | | | 230 | | | | 100 | | | |
|---|---|---|---|---|---|---|---|---|---|---|---|---|
| | Pyramids | | Octahedra | | Pyramids | | Octahedra | | Pyramids | | Octahedra | |
| | Py1 | Py2 | Oc1 | Oc2 | Py1 | Py2 | Oc1 | Oc2 | Py1 | Py2 | Oc1 | Oc2 |
| O1 | 1.85(3) | 1.98(3) | – | – | 1.88(3) | 1.97(3) | – | – | 2.02(2) | 1.87(2) | – | – |
| O2 | – | – | 1.91(4) | 1.89(4) | – | – | 2.00(3) | 1.77(3) | – | – | 1.85(4) | 1.89(5) |
| O3 | – | – | 1.84(4) | 1.88(4) | – | – | 1.75(3) | 2.00(3) | – | – | 1.91(5) | 1.86(5) |
| O4 | 1.96(2) | 1.98(2) | – | – | 1.97(2) | 1.96(2) | – | – | 1.97(2) | 1.95(2) | – | – |
| O5 | – | – | 1.96(2) | 1.92(2) | – | – | 1.96(1) | 1.92(2) | – | – | 1.91(2) | 1.97(2) |
| O6 | 2.00(1) | 1.84(1) | 1.95(1) | 2.13(1) | 1.99(1) | 1.84(1) | 1.94(1) | 2.15(1) | 1.83(1) | 2.00(1) | 2.14(1) | 1.94(1) |
| $|M|$ | 0.46(9) | 1.65(9) | 0.36(9) | 0.39(9) | 0.28(8) | 1.22(8) | 1.39(8) | 1.52(8) | 1.76(6) | 1.76(6) | 3.41(8) | 1.47(8) |

Having the most reliable information on the spin state of all the $Co^{3+}$ ions at $T = 100$ K, we try to describe in Fig. 10 the picture of orbital ordering on the basis of the ligand displacements from their average positions.[24] Unfortunately, only the displacemets $\Delta y$(O61) = −0.15(1) Å and, with a lower reliability, $\Delta y$(O62) = 0.03(1) Å are statistically significant. They are shown by the arrows in Fig. 10. The other ligand displacements are $\Delta z$(O1) = ±0.07(4) Å, $\Delta x$(O4) = ±0.01(2) Å and $\Delta z$(O2) = ±0.02(6) Å, $\Delta x$(O5) = ±0.03(2) Å in the planes of pyramids and octahedra, respectively. The $t_{2g}$ orbitals, which are close to spherical symmetry, are described by circles shadowed depending of the number of electrons transferred to $e_g$ ones.

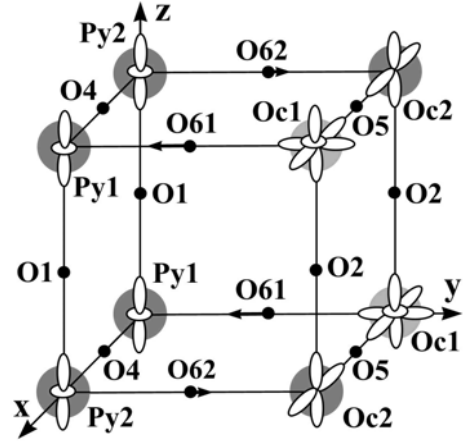

FIG. 10. Asymmetric part of the unit cell $2a_p \times 2a_p \times 4a_p$ with non-equivalent pyramidal (Py1, Py2) and octahedral (Oc1, Oc2) $Co^{3+}$ ions. The ligands ($O^{2-}$ ions O1, O2, O4, O5, O61, O62) are described by small black circles. The unit cell origin is at $x = −1/4$, $y = 0$, $z = −z_{Py1}$.

Knowing that the $Co^{3+}$ ions in the Oc1 and Oc2 sites are in the HS state and in the IS state, respectively, and there is no shift of O5 from its mean position, one may assign to the Oc2 ions the orbital $x^2 − z^2$. If there would be no shift of the O62 ligand the same orbital should be assigned to the Py2 $Co^{3+}$ ion. Taking into account small displacement of O62 towards Oc2, we chose for the Py2 the $3z^2 − r^2$ orbital, which has a higher dimension along **y** than the $x^2 − z^2$ and provides a small shift of O62 towards Oc2. Since O4 is located very precisely in the middle between Py1 and Py2, they should have the same orbital $3z^2 − r^2$.

Only in the phase 1 each spin has six opposite nearest neighbors. In two other phases exchange interaction of the nearest neighbors is either negative or positive. The sign of interaction hardly can be predicted using the Googenough-Kanamori rules[26] for the variety of the orbital configurations observed.

## VI. CONCLUSION

Discovered polymorphism that depends on the perfection of oxygen ordering among the apical positions at very close oxygen content makes very difficult investigation of magnetic structure and of its temperature evolution. With one sample containing at 308 K a single phase $2a_p \times 2a_p \times 2a_p$, $Pmma$ (Z = 2) and another sample with 75.7(1.1)% of the same phase and 24.3(6)% of second phase $a_p \times a_p \times 2a_p$, $Pmmm$ (Z = 1) we have been able to determine for the phase of interest a common in both samples magnetic structure and its temperature transformation.

Three magnetic phases have been observed for both samples: phase 1 (255 K < T < 290 K), phase 2 (170 K < T < 255 K), and phase 3 (T < 170 K). It is shown that the crystal structure of the phases 1 and 2 belongs to the space group $Pmma$ and has the unit cell $2a_p \times 2a_p \times 2a_p$. A structure phase transition with the change of the unit cell to $2a_p \times 2a_p \times 4a_p$, as well as of the symmetry to $Pcca$, takes place at $T_2 \approx 170$ K.

The wave vector of magnetic structure has been found as $\mathbf{k}_{19} = 0$ for the phases 1 and 3, while it is $\mathbf{k}_{22} = \mathbf{b}_3/2$ for the phase 2. The basis function of irreducible representations of the group $G_\mathbf{k}$ has been obtained for each phase. It is shown that the spontaneous moment in the phase 1, just below the Néel temperature, cannot be due to the spin canting of the Dzyaloshinskii-Moriya type, which is forbidden by symmetry. A spontaneous moment $M_0 = 0.30(3)$ $\mu_B$/Co is found to be due to ferrimagnetism, i.e., antiparallel ordering of the spins in two non-equivalent pyramidal sites.

Apparently, antiparallel orientation of equal moments in non-equivalent octahedral (phase 2) as well as in pyramidal (phase 3) positions stimulates metamagnetic transitions in these antiferromagnetic phases.

The moment values and the ion-ligand distances determined from the neutron diffraction experiment indicate possible spin-state/orbital ordering in each of three phases. The spin-state ordering is determined for all magnetic phases, and the picture of orbital ordering is given for the low-temperature phase. The experiments on single crystals would allow determination of the orbital ordering at all temperatures.

We suppose that all these results, being reproducible for two different samples, one of which contains considerable amount of erroneous phase, reflect actual spin/spin-state/orbital ordering in TbBaCo$_2$O$_{5.5}$.

The experiments have been performed at the Swiss Spallation Neutron Source, Paul Scherrer Institute, Villigen, Switzerland. We thank L. Keller, V. Pomjakushin and D. Sheptyakov for their help during the experiments on DMC and HRPT diffractometers. In particular, we would like to point out very useful critical comments on the manuscript made by D. Sheptyakov. This work was carried out in framework of the INTAS project (Grant N 01-0278). We are deeply indebted to INTAS for the Grant, which has made possible this study. We are grateful to the project coordinator Prof. A. Furrer for the fruitful scientific discussions and excellent management. Partial supports from the Russian Foundations for Fundamental Researches (Project N° 02-02-16981), by the grant SS-1671.20032, and by the NATO grant PST CLG 979369 are also acknowledged.